\documentclass[aps,epsf,preprint,prl,showpacs]{revtex4-1}
\usepackage[dvips]{graphicx}
\begin{document}
\bibliographystyle{apsrev}
\title{Nonlinear Optical Rectennas}

\author{A. Stolz, J. Berthelot,
L. Markey, G. Colas des Francs, and A. Bouhelier$^{*}$}
\affiliation{Laboratoire Interdisciplinaire Carnot de
Bourgogne, CNRS-UMR 6303, Universit\'{e} de Bourgogne, 21078 Dijon,
France}

\begin{abstract}
We introduce strongly-coupled optical gap antennas to interface optical radiation with current-carrying electrons at the nanoscale. The transducer relies on the nonlinear optical and electrical properties of an optical antenna operating in the tunneling regime. We discuss the underlying physical mechanisms controlling the conversion and demonstrate that a two-wire optical antenna can provide advanced optoelectronic functionalities beyond tailoring the electromagnetic response of a single emitter.  Interfacing an electronic command layer with a nanoscale optical device may thus be facilitated by the optical rectennas discussed here. 
\end{abstract}

\pacs{73.40.Jn, 73.40.Ei, 73.63.Rt, 42.65.Ky	}

 \maketitle
Giant enhancement of optical fields are generally locally produced in the feedgap formed between two or several adjacent resonant metal nano-antennas~\cite{SchullerNM10,HalasReview}. Strongly-coupled optical antennas are thus essential for amplifying weak optical interaction cross-sections such as the vibrational responses of a single molecule~\cite{huser04,gordonNL12} or nonlinear $ \chi^2$ and $\chi^3$ processes~\cite{danckwerts07,quidant08,Slablab:12,Berthelot:12OPEX}. When the separation distance between the individual constituents of the optical antenna reduces, large Coulomb splitting of the underlying plasmonic modes takes place~\cite{aizpurua06,natelson11prb,HechtNL12b}. For an atomic-scale gap, charge transfer plasmons are driving the antenna response in the so-called quantum tunneling regime~\cite{aizpuruaNC12,baumberg12,dionneNL13,LiACS13}. In this regime, the quantum nature of the interaction opens a new paradigm for utilizing optical gap antennas beyond the control of electromagnetic fields at the nanometer length scale~\cite{novotny11NP,Knight11}. For instance, incoming photons can exchange energy with tunneling charges modifying thus the conductance of the barrier~\cite{gustafson76,Cutler12} and strong-field effects were recently reported~\cite{borisovNL12,dombiNL13}.  Therefore, adopting metal-based optical antennas as a disruptive technological vehicle may provide advanced functional devices to interface nanoscale electronics and photonics. Several steps were recently made in this direction by electrically wiring optical feedgaps where an enhanced optical field is self-aligned with a large static electric field (10$^7$-10$^9$~V/m)~\cite{ward10,HechtNL12,Berthelot:12OPEX}. 

Inspired by these advances, we demonstrate here that ultrafast laser pulses can interact with tunneling charges in an electrically wired optical feedgap to create a nanometer-scale nonlinear rectifying antenna, or rectenna, operating at optical frequencies. We show that the rectenna's nonlinear transducing yield is driven by the electrical environment of the feedgap. Specifically, the crossover between a linear intensity dependence of the rectification yield to a multi-order power law is strongly reduced at the onset of Fowler-Nordheim tunneling. 

Nanometer-scale optical gap antennas are produced by controlling the electromigration of 100~nm wide 5~${\rm\mu}$m long gold nanowires fabricating on a glass substrate~\cite{mceuen99,Mangin2009}. The nanowires and their electrical connections are realized by a double-step lithography. Macroscopic electrodes and a series of alignment marks are first fabricated by standard optical lithography. The marks are used for subsequently defining nanowires by electron-beam lithography.  A 2~nm thick layer of Cr followed by  50~nm of Au are then evaporated to form the metal structures. A liftoff of the electron-sensitive resist finalizes the sample. 
\begin{figure}[t]
\includegraphics[width=9cm]{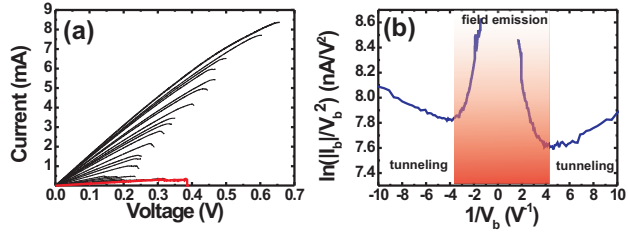}
\caption{\label{figure1} (Color online). (a) Current-voltage plot of a Au nanowire for a series of subsequent bias sweeps. For each sweep, the voltage ramp is manually stopped before electrical failure of the nanowire. The resistance increases between sweeps as the result of the formation of a constriction. Eventually, the process leads to the electrical failure of the nanowire (red curve).  (b) Fowler-Nordheim representation of the electrical characteristic of an optical gap antenna produced by electromigration. The regime of transport (direct tunneling or field emission) is dictated by the applied bias  $V_{\rm b}$.}
\end{figure}

\begin{figure*}[t]
\includegraphics[width=18cm]{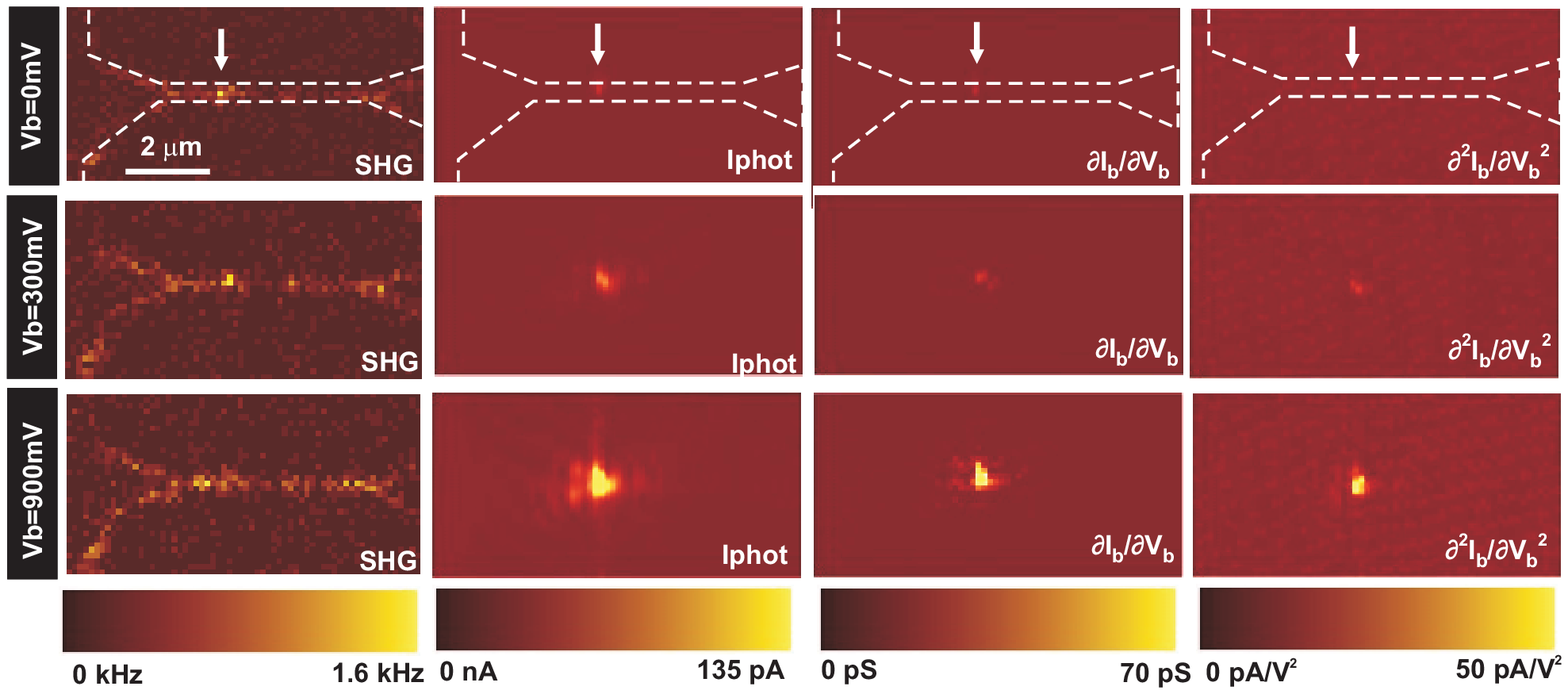}
\caption{\label{figure2} (Color online).  ($x,y$) spatial dependence of the second harmonic signal (first column on the left), the demodulated laser-induced current $I_{\rm phot}$ (second column), the differential conductance $\partial I_{\rm b}/\partial V_{\rm b}$ (third column) and the nonlinear conductance $\partial^{2} I_{\rm b}/\partial V_{\rm b}^{2}$ (last column) for three biases $V_{\rm b}$=0~mV, $V_{\rm b}$=+300~mV, and $V_{\rm b}$=+900~mV, respectively. The electrode and nanowire contours are represented by the dashed layout for $V_{\rm b}$=0~mV. The arrows show the spatial location of the antenna feedgap. The bias is applied to the left electrode.}
\end{figure*}

Figure~\ref{figure1}(a) shows a typical current-voltage plot of a contacted nanowire when a ramping bias is applied to the terminals. At the onset of electromigration, the curve is departing from an Ohmic behavior~\citep{StahlmeckeAPL06}. The formation of a constriction preceding that of a nanometer-scale gap  is controlled by applying successive bias sweeps that eventually lead to the electrical breakdown of the nanowire (red curve)~\cite{Mangin2009}. We find that approximately 60\% of the junctions produced by this method lead to gap sizes commensurate with electron tunneling. The Folwer-Nordheim representation of Fig.~\ref{figure1}(b) illustrates the current response after electromigration.  The current $I_{\rm b}$ is measured with a $10^9$~V/A current-to-voltage converter. The minima in the curve at -3.8~V$^{-1}$ and 4.2~V$^{-1}$ indicate the cross-over from direct-tunneling transport to a field emission regime~\cite{Mangin2009}. The height $\phi$ of the potential barriers can be directly deduced from these two minima~\cite{Beebe06}. We measure $\phi^{\rm b}$=0.23~eV and $\phi^{\rm g}$=0.26~eV for the barriers at the biased and ground electrodes, respectively. These barrier heights are consistent with reports of reduced effective work functions in Au electromigrated gaps operating in the presence of adsorbates~\cite{steinmann:3178,Mangin2009}.  Although constituted from the same material, the slightly different barrier heights are linked to geometrical asymmetry of the gap~\cite{mayer2011}, a typical characteristic of electromigrated junctions~\cite{Berthelot:12OPEX}. Asymmetry in the Fowler-Nordheim representation was systematically observed in our experiments. When measurable, crossover between direct-tunneling and field emission was also consistently observed at $\phi<$0.5~eV.

We then interrogate the nonlinear rectifying properties of such contacted optical gap antenna by laterally scanning antenna through the focus of a tightly focused Titanium:Sapphire pulsed laser beam. The photon energy is fixed at 1.53~eV. The pulse duration at the exit of the laser and the repetition rate are 180~fs and 80~MHz, respectively. The numerical aperture (N.~A.) of the focusing objective is 1.49. For each lateral ($x,y$) position of the antenna in the focus, we conduct photon-dependent inelastic tunneling spectroscopy by measuring the first and second derivatives of the current. To that purpose, a small modulated bias $V_{\rm mod} = V_{\rm pp} \cos(2\pi F_{\rm mod}t)$  is added to the static voltage $V_{\rm b}$ applied to the nanojunction with $V_{\rm pp}$=30~mV and $F_{\rm mod}$=1~kHz.  The differential conductance $\partial I_{\rm b}/\partial V_{\rm b}$ of the gap antenna and its nonlinearity $\partial^{2} I_{\rm b}/\partial V_{\rm b}^{2}$ are  measured by two lock-in amplifiers referenced at $F_{\rm mod}$ and $2F_{\rm mod}$, respectively.  The laser-induced current $I_{\rm phot}$ is measured by chopping the laser beam at $F_{\rm chp}$=400~Hz and demodulating the tunneling current at $F_{\rm chp}$ with a third lock-in amplifier. Finally, we also record the second-harmonic generation (SHG) activity of the gold structure by detecting the harmonic radiation emitted at 3.06~eV with a single-photon avalanche photodiode. Due to the geometrical asymmetry of the feedgap, the tunneling junction exhibits an enhanced SHG response~\cite{Berthelot:12OPEX} providing an imaging contrast to locate the antenna feedgap.

Figure~\ref{figure2} displays a series of lateral scans of the different gap responses under a constant average laser intensity at focal spot of $840$~kW/cm$^2$.  The polarization of the laser is aligned along the nanowire for the remaining of the discussion. The simultaneously measured SHG,  photocurrent $I_{\rm phot}$,  linear conductance $\partial I_{\rm b}/\partial V_{\rm b}$ and nonlinear conductance $\partial^{2} I_{\rm b}/\partial V_{\rm b}^{2}$ are displayed for a null bias and for $V_{\rm b}$=+300~mV, and $V_{\rm b}$=+900~mV, respectively. The dashed lines in the upper images of Fig.~\ref{figure2} ($V_{\rm b}$=0~V) follows the contours of the two side electrodes and that of the nanowire. While the SHG signal is expectedly large at the edges of the Au structure, the strongest nonlinear response is observed at the position of the electromigrated gap~\cite{Berthelot:12OPEX} as indicated by the arrows. When the femtosecond-pulsed laser overlaps the feedgap,  a 20~pA photocurrent is produced at $F_{\rm chp}$ in the absence of an applied bias ($V_{\rm b}$=0~V).  At the same position,  electrical changes of the first and second derivatives of the antenna conductance $\partial I_{\rm b}/\partial V_{\rm b}$ and $\partial^{2} I_{\rm b}/\partial V_{\rm b}^{2}$ are measured. The concomitant occurrence of these signals at a very well-defined spot strongly support that rectification is the dominating mechanism. 

Thermal expansion of the electrodes upon photon absorption would also effectively change the conductance; this was extensively debated in the context of photo-assisted transport in scanning tunneling microscopy~\cite{grastrom02}. However, thermal expansion was shown to be negligible in planar metal junctions lying on a substrate and illuminated under a constant-wave (CW) laser excitation~\cite{guhr07,ward10}. For a pulsed excitation, the effect of a light-induced thermal expansion in the $I_{\rm phot}$ or $\partial I_{\rm b}/\partial V_{\rm b}$ maps of Fig.~\ref{figure2}  would be distributed along the nanowire and not be uniquely restricted to the rectenna feedgap. We occasionally observed a weak photocurrent produced by thermally-excited electrons with a typical sign reversal when the laser illuminates either side of the metal rectenna (see Fig.~S1 of the Supplemental Material~\cite{supplmat}).  

The yield of optical rectification is generally driven by the classical responsivity of the device $S=~[\partial^{2} I_{\rm b}/\partial V_{\rm b}^{2}][\partial I_{\rm b}/\partial V_{\rm b}]^{-1}$ and therefore strongly depends on the electrical nonlinearity of the barrier~\cite{mayer2011}. To verify that the device discussed here is behaving semi-classically, we operated the rectenna at different point of the electrical characteristic by applying a static voltage $V_{\rm b}$ across the feedgap. The lateral responses of the rectenna for  $V_{\rm b}$=+300~mV and  $V_{\rm b}$=+900~mV are shown in the series of images in the middle and bottom lines of Fig.~\ref{figure2}. For this particular device and under these voltages, electron transport in the rectenna is driven by direct-tunneling (data not shown). The differential conductance $\partial I_{\rm b}/\partial V_{\rm b}$ of the rectenna increases with bias because of the growing number of electrons injected in the feedgap. In this region of the characteristic, the nonlinearity of the conductance $\partial^{2} I_{\rm b}/\partial V_{\rm b}^{2}$ becomes larger as demonstrated in Fig.~\ref{figure2}. The classical responsivity
at the feedgap is $S$~(0~mV)= 0.2~V$^{-1}$ and increases linearly with bias with a slope of $\sim$0.3~V$^{-2}$. We measure a six-fold enhancement of $I_{\rm phot}$ when $V_{\rm b}$ is increased from 0~mV to +900~mV. 
The simultaneously recorded SHG signal is stable within a few percent during the bias increments. The variation is attributed to a slight drift of the focus between scans. The relative stability of the harmonic response is a good indication that the structural integrity of the rectenna is maintained during the experiment: SHG is strongly affected by symmetry and local defects, and any significant modifications of the feedgap geometry or the leads would have been recorded in the SHG maps.

\begin{figure}[t]
\includegraphics[width=9cm]{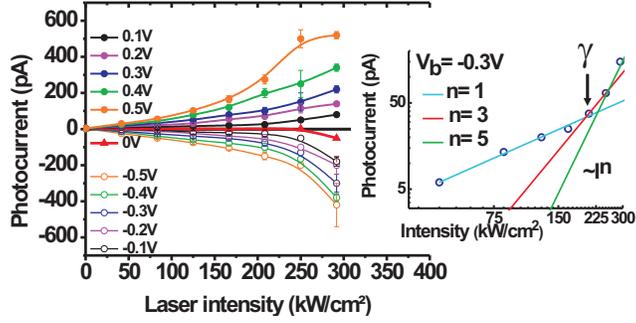}
\caption{\label{figure3} (Color online). Dependence of the photocurrent $I_{\rm phot}$ on the average laser intensity for different symmetric biases $V_{\rm b}$. Inset: Logarithmic plot of $I_{\rm phot}$ versus average laser power for $V_{\rm b}$=-0.3~V. For moderate pulsed laser intensities the photocurrent scales linearly with power. Higher-order photon processes are occurring when the laser intensity is increased. $\gamma$ denotes the transition from a linear regime to a power law dependence with $n$~$>1$.}
\end{figure}

Figure~\ref{figure3}  shows the evolution of the rectified current $I_{\rm phot}$ with the average laser intensity for a series of negative and positive voltages $V_{\rm b}$. The data were acquired from a different rectenna. The laser fluence dictates the regime of operation of the optical rectenna.  For moderate laser intensities and regardless of the applied bias, the rectenna operates in a linear regime. This regime consistent with the behavior reported in Ref.~\cite{Ostrikov12} for a discontinuous gold film. An example of a linear relationship is shown in the double-logarithmic plot in the inset of Fig.~\ref{figure3}  for $V_{\rm b}$~=~-300~mV. The data below $\sim$200~kW/cm$^2$ ($\gamma$ point) are well fitted by a linear trend with a slope of 14~pA/kWcm$^{-2}$.  At $\gamma$, we estimate a conversion of 4$\times$10$^{-4}$ electron per received photon by the rectenna feedgap assuming a sensitive area of 1~nm$\times$100~nm. After the threshold $\gamma$, the trend is no longer linear; multiphoton processes are contributing to the photocurrent as indicated by the increasing power-law exponent with laser intensity. We rule out a photo-induced thermionic emission of carriers to explain the nonlinear trend. For a fixed sign of the bias and the relatively small asymmetry of the junction, a thermal excitation of electrons across the barrier would provide a contrast reversal between the two sides of the rectenna's feedgap. Multiphoton absorption leads to above-threshold photoemission for excitation energies above the barrier~\cite{Hommelhoff10} or to tunneling of excited electrons in states located above the Fermi surface but remaining below the work function~\cite{Diesing04}.  The electrical conductance in a nanoscale junction typically involves tunneling of electrons located near the Fermi level $\varepsilon_{\rm F}$, which for gold are located within the $sp$ conduction band. Taking into account the averaged reduced effective work function of the electrodes at rectenna feedgap, a single photon absorption of an electron located near $\varepsilon_{\rm F}$ has an excess energy of $\sim$1.2~eV above the effective barrier. Under this excitation condition, the photocurrent resulting from a one-photon or even a multiphoton above-threshold emission should be independent of the classical responsivity $S$ of the rectenna. This is in direct contrast with Fig.~\ref{figure2} and Fig.~S1 where even a small bias significantly increases the rectenna's photoconductance. We verify with Fig.~S2 of the Supplemental Material~\cite{supplmat} that the rectification picture remains valid even for high laser intensities by correlating the bias dependence of the photocurrent $I_{\rm phot}$ with the nonlinear conductance $\partial^{2} I_{\rm b}/\partial V_{\rm b}^{2}$ for the range of laser intensities used in this work. 

J.~K.~Viljas and J.~C.~Cuevas calculated that the photoconductance of a contact between two Au atoms increases when $d$-band electrons are optically excited~\cite{Cuevas07}. Transition of $d$ electrons to $\varepsilon_{\rm F}$ occurs for optical energies approximately comprised between 2.5~eV and 7.5~eV~\cite{Eckard84}, which for our laser, requires a multiphoton absorption process with a minimum order $n\geq$~2. This nonlinear absorption is rather efficient for gold nanostructures~\cite{bouhelier03b,dulkeith04}, and especially optical gap antennas~\cite{quidant08}. Upon a pulsed laser excitation a strong nonlinear photoluminescence response is typically observed resulting from a radiative interband recombination of an $sp$ electron promoted above  $\varepsilon_{\rm F}$ with a hole in the $d$ band~\cite{Imura05,Biagoni09}. Considering the above arguments, we hypothesize that optical excitations of $d$ band electrons to states above the Fermi level are contributing to  $I_{\rm phot}$ and are responsible for the nonlinearity of the rectenna displayed in Fig.~\ref{figure3}. Since $I_{\rm phot}$ and $\partial^{2} I_{\rm b}/\partial V_{\rm b}^{2}$ follow the same bias evolution (Fig.~S2), the higher order power dependence shown in the inset of Fig.~\ref{figure3} suggests that optically excited low-lying electrons are contributing to the photocurrent. 

\begin{figure}[t]
\includegraphics[width=9cm]{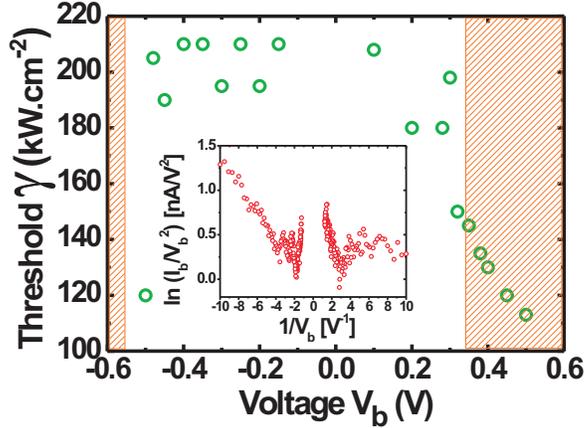}
\caption{\label{figure4} (Color online). Evolution of the threshold $\gamma$ defining the transition between the rectenna's linear to nonlinear regime as a function of bias $V_{\rm b}$. The shaded areas show the voltage ranges in which the rectenna is in an electrical field emission regime. The inset shows the Folwer-Nordheim representation of the rectenna's electrical characteristic. The two minima at -1.8~V$^{-1}$ and 2.8~V$^{-1}$ are the cross over biases from direct tunneling to field emission tunneling.}
\end{figure}

We verify this hypothesis by plotting the evolution with bias of the threshold value $\gamma$ delimiting the linear to the nonlinear regime as defined in the inset of Fig.~\ref{figure3}. $\gamma$ represents the laser intensity from which $d$-band electrons need to be taken into account in the photoconductance. The bias dependence of the $\gamma$ threshold is reported in Figure~\ref{figure4}. For -0.45~V$<V_{\rm b}<$0.3~V, $\gamma$ is fairly constant at around 200~$\pm$11~kW/cm$^2$. In this range of voltage the rectenna operates in a direct tunneling regime ($eV_{\rm b}<\phi^{\rm b}$ and $\phi^{\rm g}$). The Fowler-Nordheim plot of the rectenna's electrical characteristic is reported in the inset of Fig.~\ref{figure4}. Here $\phi^{\rm b}$=0.55~eV and $\phi^{\rm g}$=0.35~eV. For larger $V_{\rm b}$, we observe a sudden drop ($\sim 40\%$) of the nonlinear threshold $\gamma$. This step-like decrease occurs at the onset of field emission as indicated by the shaded areas in Fig.~\ref{figure4}. The figure indicates that for a laser intensity near the $\gamma$ point at low biases, a static reduction of the barrier height increases the number of conduction channels opened by multiphoton absorption of $d$ band electrons. This bias sensitivity provides thus a handle to control the rectenna's nonlinear characteristic. 

To conclude we show here that gaps formed by electromigrating nanowires can act as an elementary nonlinear rectifier when irradiated by a femtosecond pulsed laser. Although the rectenna's geometry is relatively simple, the optical gap and the electrical rectifying gap are self-aligned at the nanometer scale. Enhancement of the rectification yield can be achieved by controlling the intrinsic feed characteristics of these two overlapping functional gaps. Engineering the barrier height and the geometrical asymmetry of the facing edges will increase substantially the classical responsivity $S$ of the rectenna~\cite{mayer2011}. Deploying a resonant plasmonic feedgap will improve the interaction cross-section of the rectenna with the incoming photons~\cite{Mayer09,HechtNL12} and will provide an enhancement of the localized electromagnetic field~\cite{natelson11prb}. These combined effects will contribute to increase the efficiency of the device. 
Finally, we note that Fig.~\ref{figure2} and Fig.~\ref{figure3} show time-averaged photocurrent maps and laser intensity dependences. Because the laser produces 180~fs short pulses at a repetition rate of 80~MHz, the measured current at the terminals is constituted of a rectified electron packets bunched on a similar time scale. By using pulse-picking technique, femtosecond single electron pulses can thus be delivered on-demand in the circuitry. This simple optical rectenna could therefore be implemented as a wired ultrafast electron source for studying temporal dynamics of nanoscale electronic devices and serve as an extremely local source of terahertz radiation.

A.~B. thanks valuable discussions with A. G. Borisov, A. Dereux and V. Meunier. The research leading to these results has received funding from the European Research Council under the European Community's Seventh Framework Program FP7/2007-2013 Grant Agreement no 306772. This project is in cooperation with the Labex ACTION program (contract ANR-11-LABX-01-01). A.~S. acknowledges a scholarship from  R\'egion de Bourgogne under the PARI initiative.

$^*$Corresponding author: alexandre.bouhelier@u-bourgogne.fr

\end{document}


\bibliographystyle{apsrev}

\title{Supplementary information for ``Nonlinear Optical Rectennas''}
\author{A. Stolz, J. Berthelot,
L. Markey, G. Colas des Francs, and A. Bouhelier$^{*}$}
\affiliation{Laboratoire Interdisciplinaire Carnot de
Bourgogne, CNRS-UMR 6303, Universit\'{e} de Bourgogne, 21078 Dijon,
France}

\begin{abstract}
This supplementary material provides additional details about thermal effects observed in optical rectennas, and describes the observed correlation between the photocurrent and the nonlinearity of the conductance when irradiated with increasing laser intensities.
\end{abstract}

 \maketitle

\section{Thermal effects}
\begin{figure*}[tbh]
\includegraphics[width=14cm]{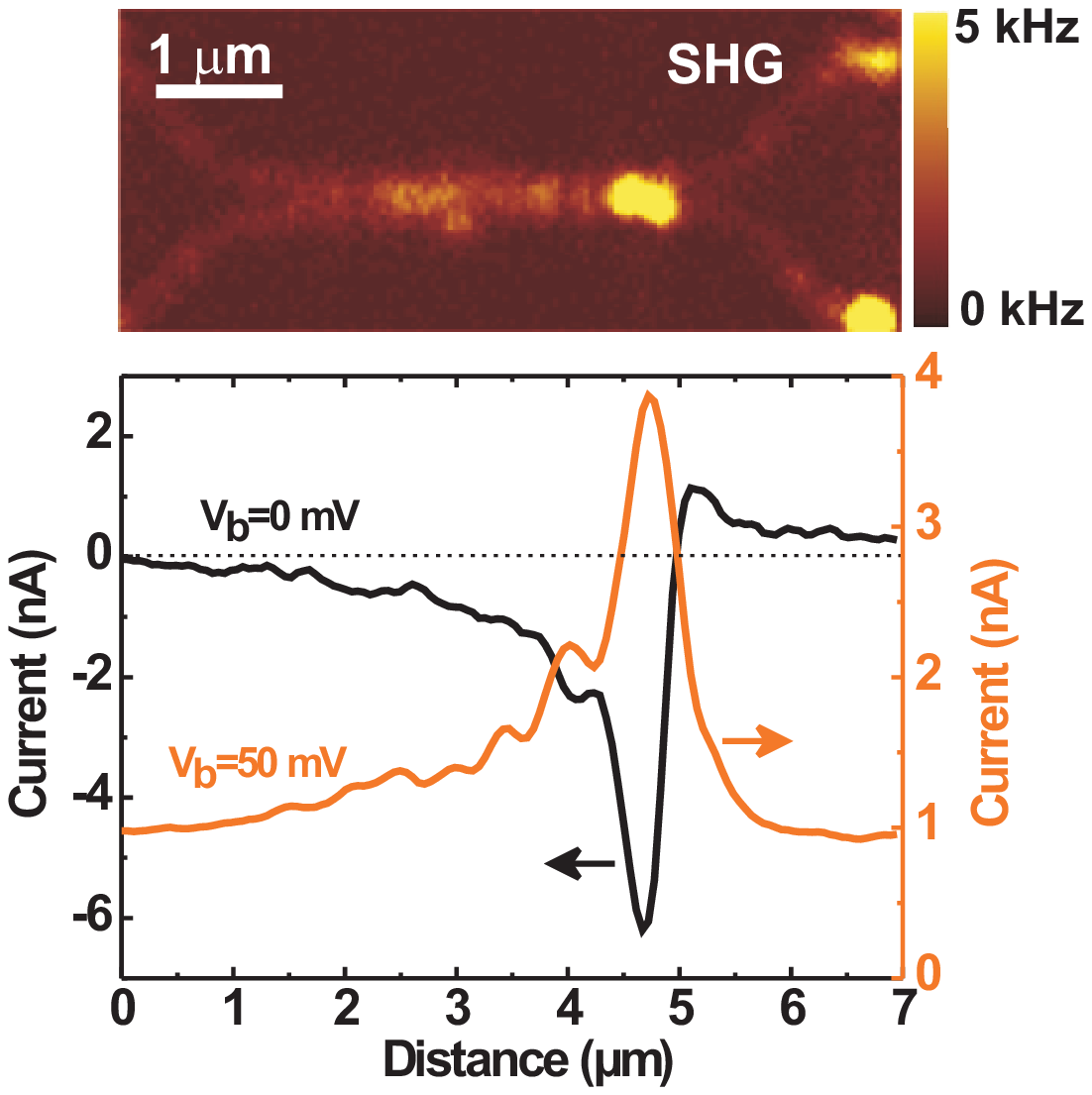}
\caption{\label{S1} (Supplemental). Confocal maps showing the second-harmonic signal enhanced at the rectenna feedgap. The graph displays the current flowing through the rectenna for $V_{\rm b}$=0~mV and 50~mV. The laser intensity is kept constant at an average value of 4~MW/cm$^{2}$. Note the different current scales in the graph.}
\end{figure*}

Thermal effects can drastically influence the conduction properties of a laser irradiated tunneling jonction as reviewed by S. Grafstr\"om in Ref.~\cite{grastrom02}. Among possible processes resulting from photon absorption in the metal and subsequent Joule heating, let us mention thermal expansion of the gold leads, thermally-assisted field emission and thermovoltage arising from a temperature difference of the two sides of the rectenna feedgap when the laser is irradiating either one~\citep{kopp12}. While it is difficult to completely rule out a potential heat-related increased conductance in the results shown in the main manuscript, heat processes are providing well defined characteristics when the sample is scanned through the focus such as a contrast reversal of the current~\cite{mceuen11}. We did observe evidence for thermal currents in a rectenna consisted of an electromigrated Ti/Au stack. The $\sim$5~nm thick Ti layer acts a seed layer between the glass substrate and the gold layer. In the main manuscript, a Cr layer was systematically used instead of Ti and the results discussed below were never reproduced with Cr/Au rectennas. Figure~\ref{S1} shows an example of a thermal effect. In this experiment, the laser intensity was intentionally kept large at 4~MW/cm$^{2}$. The confocal map of Fig.~\ref{S1} displays the spatial dependency of the second harmonic generation of the structure. One recognizes the outline of the nanowire and the contacting electrodes at the left and right of the image. The left electrode is at ground potential. At the location of the rectenna feedgap near the right contact, the second harmonic activity is amplified due to the irregular geometry of the gap and a possible electromagnetic field enhancement~\cite{Berthelot:12OPEX}. The graph of Fig.~\ref{S1} displays the position dependence of the current flowing through the rectenna for null bias and $V_{\rm b}$=50~mV taken along the nanowire. At zero bias, there is a clear change of the current sign between the two sides of the feedgap. The influence of the laser absorption can also be traced along the left part of the nanowire where the negative value of current increases to reach its maximum at the rectenna's feedgap. Because the rectenna was produced near the top electrode, dissipation is therefore reduced compared to the side located the closest to the macroscopic heat sink (right side). When the rectenna is biased, thermally-excited electrons are no longer the predominant transport channel. The current response of the rectenna when $V_{\rm b}$ is increased to 50~mV demonstrates that even at a very small bias, the ($x,y$) response of the current is symmetric with respect to the feedgap and the current magnitude is increased as shown by the different scales. The increased current with $V_{\rm b}$ can be understood in term of the classic responsivity $S$ discussed in the main text. 

\section{Optical rectification}
Within the assumptions discussed by Mayer and co-workers in Ref.~\cite{mayer2011} and by Ward \textit{et al}. in Ref.~\cite{ward10}, the classical picture of the rectified current is given by 
\begin{equation}
\label{rectification}
I= I_{\rm b}+I_{\rm phot}= I_{\rm b}+\frac{V_{\rm opt}^2}{4}\frac{\partial^{2} I_{\rm b}}{\partial V_{\rm b}^{2}}.
\end{equation}

\begin{figure*}[h]
\includegraphics[width=16cm]{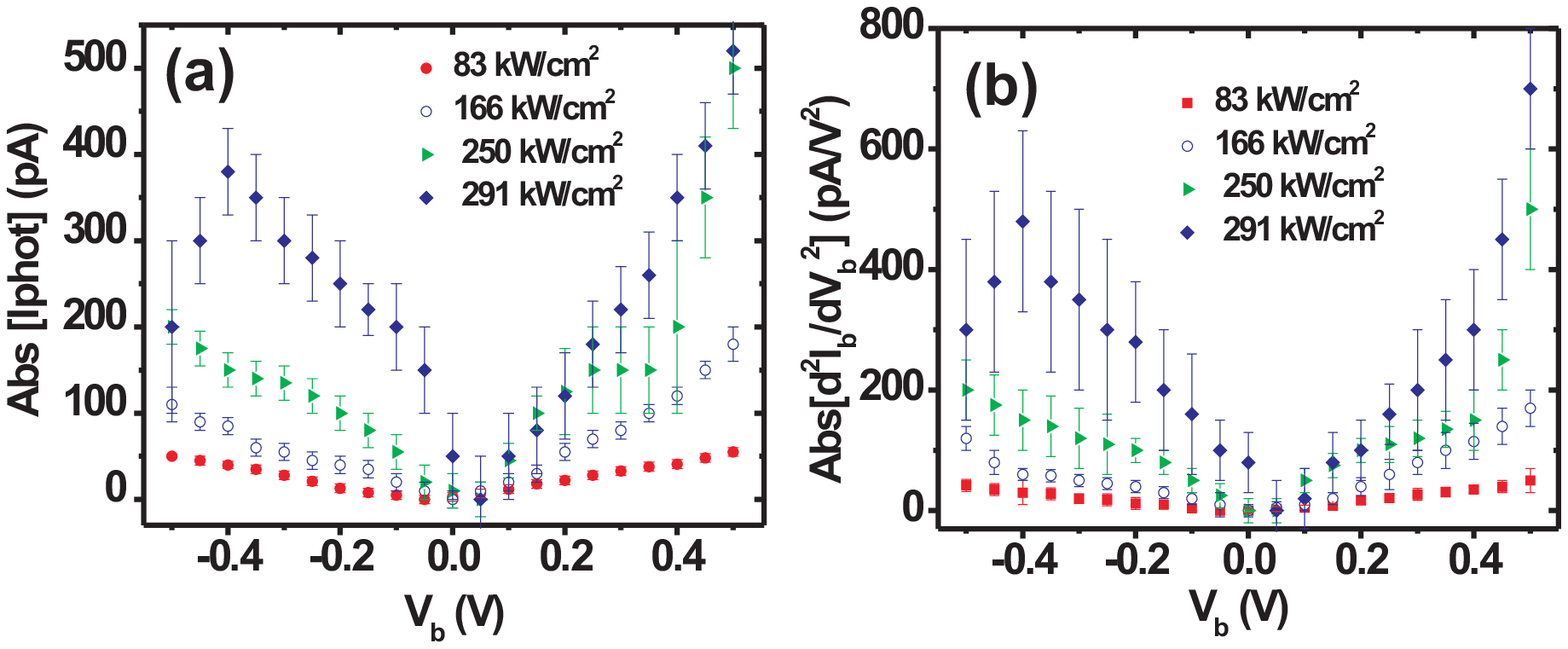}
\caption{\label{S2} (supplemental). Bias dependence of the absolute value of the photocurrent (a) and the nonlinearity of the rectenna conductance (b) for increasing laser intensity. $I_{\rm phot}$  and $\partial^{2} I_{\rm b}/\partial V_{\rm b}^{2}$ are correlated indicating that the rectification picture is the dominant mechanism for all the laser intensities and bias values investigated.}
\end{figure*}

$I_{\rm b}$ is the static tunneling current of the biased rectenna and is filtered out by the lockin amplifier referenced at the chopping frequency of the laser. $V_{\rm opt}$ is the light-induced voltage created at the rectenna feedgap. Equation~\ref{rectification} indicates that the photocurrent $I_{\rm phot}$ is proportional to the optical induced voltage and the nonlinearity of the conductance $\partial^{2} I_{\rm b}/\partial V_{\rm b}^{2}$.  Figure~\ref{S2} shows the bias dependence of the photocurrent $|I_{\rm phot}|$ created by the rectenna and the nonlinearity of the conductance $|\partial^{2} I_{\rm b}/\partial V_{\rm b}^{2}|$ measured simultaneously for different laser intensities. The data here are obtained with the rectenna discussed in Fig.~3 and Fig.~4 of the main manuscript. The evolution of the photocurrent curves for all bias values and laser intensities are well reproduced by $|\partial^{2} I_{\rm b}/\partial V_{\rm b}^{2}|$. The correlation between Fig.~\ref{S2}(a) and Fig.~\ref{S2}(b) strongly suggests that optical rectification remains at play even when the rectenna is irradiated under laser intensities higher than the linear threshold $\gamma$ discussed in the main manuscript.

